\documentclass[12pt]{iopart}
\input epsf

\begin{document}

\title{Evaluation of configurational entropy of a model 
liquid from computer simulations}

\author{Srikanth Sastry}
\address{Jawaharlal Nehru Centre for Advanced Scientific Research, 
Jakkur Campus, Bangalore 560064, INDIA}

\begin{abstract}
Computer simulations have been employed in recent years to evaluate
the configurational entropy changes in model glass-forming liquids.
We consider two methods, both of which involve the calculation of the
`intra-basin' entropy as a means for obtaining the configurational
entropy. The first method involves the evaluation of the
intra-basin entropy from the vibrational frequencies of inherent
structures, by making a harmonic approximation of the local potential
energy topography. The second method employs simulations that confine
the liquid within a localized region of configuration space by the
imposition of constraints; apart from the choice of the constraints,
no further assumptions are made. We compare the configurational
entropies estimated for a model liquid (binary mixture of particles
interacting {\it via} the Lennard-Jones potential) for a range of 
temperatures, at fixed density. 
\end{abstract}


%\submitted{{\noindent \it \today}}

\section{Introduction}

	Whether a thermodynamic phase transition underlies the
transformation of a supercooled liquid into an amorphous solid, or
{\it glass}, at the laboratory glass transition temperature $T_g$ is
among the central questions addressed by numerous researchers studying
the supercooled liquid and glassy states. The notion of
configurational entropy\cite{gibbs-dimarzio,adam-gibbs} has played a
significant role in attempts to define and understand the
thermodynamic nature of the glass transition. In recent times, there
have been various attempts to determine the configurational entropy of
realistic liquids analytically and by computer simulations
\cite{speedy1,speedy2,speedy2b,speedy3,fs,fs2,heuer,parisi,scala,ruocco,silvio,mezard,parisi2,sastryprl}. The
purpose of this paper is to compare two such methods that have been
studied recently, namely the evaluation of the configurational entropy
{\it via} the analysis of local potential energy minima or {\it
inherent structures}
(IS)\cite{inh,fs,fs2,heuer,parisi,parisi2,sastryprl}, and by the
calculation of {\it basin} free energies by confining the liquid
within a localized region of configuration space by the imposition of
constraints\cite{speedy3}. These approaches, and results from their
implementation, are described in the following sections.

The model liquid studied is a binary mixture of $204$ type $A$ and
$52$ type $B$ particles, interacting {\it via} the Lennard-Jones (LJ)
potential, with parameters $\epsilon_{AB}/\epsilon_{AA} = 1.5$,
$\epsilon_{BB}/\epsilon_{AA} = 0.5$, $\sigma_{AB}/\sigma_{AA} = 0.8$,
and $\sigma_{BB}/\sigma_{AA} = 0.88$, and $m_B/m_A = 1$, which has
been extensively studied as a model glass
former\cite{kob,sastry,fs,parisi,sastryprl}. Results presented in Sec. II are
from molecular dynamics simulations, at a reduced density $\rho =
1.2$, which have been described in detail elsewhere\cite{sastry,sastryprl}.
Since the density is fixed, the dependence on density is not always
shown explicitly in the following.

\section{Configurational Entropy from Inherent Structures}

	In the inherent structure approach\cite{inh}, one considers
the division of configurational space into basins of local potential
energy minima.  In practice such basins may be defined as the set of
all points in configurational space that maps to the given local
minimum under a specified local energy minimization procedure. Quite
generally, one may then write the total partition function of the
system as a sum of restricted partition function integrals over
individual basins. Rewriting the partition function in this way
introduces an entropy term associated with the number of local
potential energy minima. With the expectation that configurations
within a given basin are accessible to each other by thermal agitation
while those belonging to distinct minima may not be, the number of
distinct potential energy minima can be seen to be a measure of the
number of physically distinct configurations or structures the system
can adopt, {\it i. e.} a measure of the configurational entropy.

Thus, the canonical partition function is re-written
as a sum over all local potential energy minima, which introduces a
distribution function for the number of minima at a given energy:
\begin{eqnarray}
Q_N(\rho,T) = \Lambda^{-3N} {1\over N_A! N_B!} \int d{\bf r}^N exp\left(-\beta \Phi\right)\\
	    = \sum_\alpha exp\left(-\beta \Phi_\alpha\right) \Lambda^{-3N} \int_{V_\alpha} d{\bf r}^N exp\left(-\beta(\Phi-\Phi_\alpha)\right)\nonumber \\
	    = \int d\Phi_\alpha ~ \Omega(\Phi_\alpha) ~exp\left(-\beta(\Phi_\alpha + N f_{basin}(\Phi_\alpha,T))\right)\nonumber \\
	    = \int d\Phi_\alpha ~exp\left(-\beta (\Phi_\alpha + N f_{basin}(\Phi_\alpha,T) - T S_c(\Phi_\alpha))\right)\nonumber
\end{eqnarray}
where $\Phi$ is the total potential energy of the system, $\alpha$
indexes individual inherent structures, $\Phi_\alpha$ is the potential
energy at the minimum, $V_{\alpha}$ is the basin of inherent structure
$\alpha$, $\Omega(\Phi_\alpha)$ is the number density of inherent
structures with energy $\Phi_\alpha$, and the configurational entropy
$S_c \equiv k_B \ln \Omega$ (Note that here $S_c$ is a function of 
energy; the equilibrium average of this quantity displayed in Fig. 4
as a function of temperature).

	The probability of finding the system in the basin of an inherent
structure of a given energy is given by the above as, 

\begin{equation} 
P(\Phi_\alpha, T) = {1\over Q_N(\rho,T)} exp\left(-\beta (\Phi_\alpha + N f_{basin}(\Phi_\alpha,T) - T S_c(\Phi_\alpha))\right). 
\label{scdos} 
\end{equation}

The probability distribution $P$ can be obtained from computer
simulations, and offers a means of obtaining $S_c$, provided one can
estimate $Q_N$ (equivalently the free energy $A(\rho,T)$ of the system)
and the basin free energy $f_{basin}(\Phi_\alpha,T)$. 

The free energy at any desired temperature is obtained from
thermodynamic integration of pressure and potential energy data from
MD simulations\cite{fs,speedy3}.  The absolute free energy $A(\rho,
T)$ of the system at density $\rho$ at a reference temperature $T_r =
3.0$ is first defined in terms of the ideal gas contribution
$A_{id}(\rho, T)$ and the excess free energy $A_{ex}(\rho, T)$
obtained by integrating the pressure from simulations:
\begin{eqnarray} 
A(\rho,T) = A_{id}(\rho,T) + A_{ex}(\rho,T),\\
\beta A_{id}(\rho,T)  = N \left(3~\ln \Lambda + \ln \rho - 1\right),\nonumber \\
\beta_r A_{ex}(\rho,T_r) = \beta_r A^{0}_{ex}(0,T_r) + N  
\int_{0}^{\rho} {d\rho^{'} \over \rho^{'}}\left( { \beta_r P\over \rho^{'} } - 1\right),\nonumber \\
\beta_r A^{0}_{ex}(0,T_r) = - \ln {N! \over N_A! N_B!}.\nonumber
\end{eqnarray}
Here, $N$ is the number of particles, $\beta \equiv k_B T$, $\Lambda$
is the de Broglie wavelength, and $A^0_{ex}$ arises from the mixing
entropy. $A_{ex}$ at a desired temperature may
be evaluated by integrating the potential energy, $E$:
\begin{equation} 
\beta A_{ex}(\rho, \beta) =  \beta A_{ex}(\rho,\beta_r) + 
\int_{\beta_r}^{\beta} E(\rho, \beta^{'}) d\beta^{'} 
\label{Efit}
\end{equation}
As observed in \cite{fs,parisi}, the $T$ dependence of $E$ at the
studied density is well described by the form $E(\rho, T) \sim
T^{3/5}$, in agreement with predictions for dense
liquids\cite{rosetar}. A fit of the potential energy data to this form
affords a means of extending with confidence the temperature
dependence of $E$ to $T$ values where direct MD data is unavailable.

The {\it basin free energy} $f_{basin}(\Phi_\alpha,T)$ is obtained by
a restricted partition function sum over a given inherent structure
basin, $V_\alpha$. For sufficiently low temperatures, one may expect
the basin to be harmonic to a good approximation. In the harmonic
approximation, we have
\begin{equation}
\beta f  = {3\over 2} \ln({\beta \over2\pi}) + {1 \over 2N} \sum_i^{3N-3} \ln \lambda_i \equiv \beta f_{therm} + \beta f_{vib},
\end{equation}
where $\lambda_i$ are eigenvalues of the Hessian or curvature matrix
at the minimum. For individual minima, these eigen values are obtained
by numerical diagonalization of the Hessian. The basin free energy can
then be obtained either as a function of the inherent structure energy
(by averaging free energies within individual energy bins) or as a
function of temperature, by averaging all inherent structures sampled
at a given temperature. $\beta f_{vib}$ is a slowly varying function
of temperature (the temperature dependence is obtained by averaging
over $1000$, $100$ inherent structures at $T < 1.$, $T>1.$
respectively), and is fitted to the form $\beta f_{vib}(T) = f_0 +
f_1/T^2$ which fits available data quite well.

\begin{figure} 
\hbox to\hsize{\epsfxsize=0.8\hsize\hfil\epsfbox{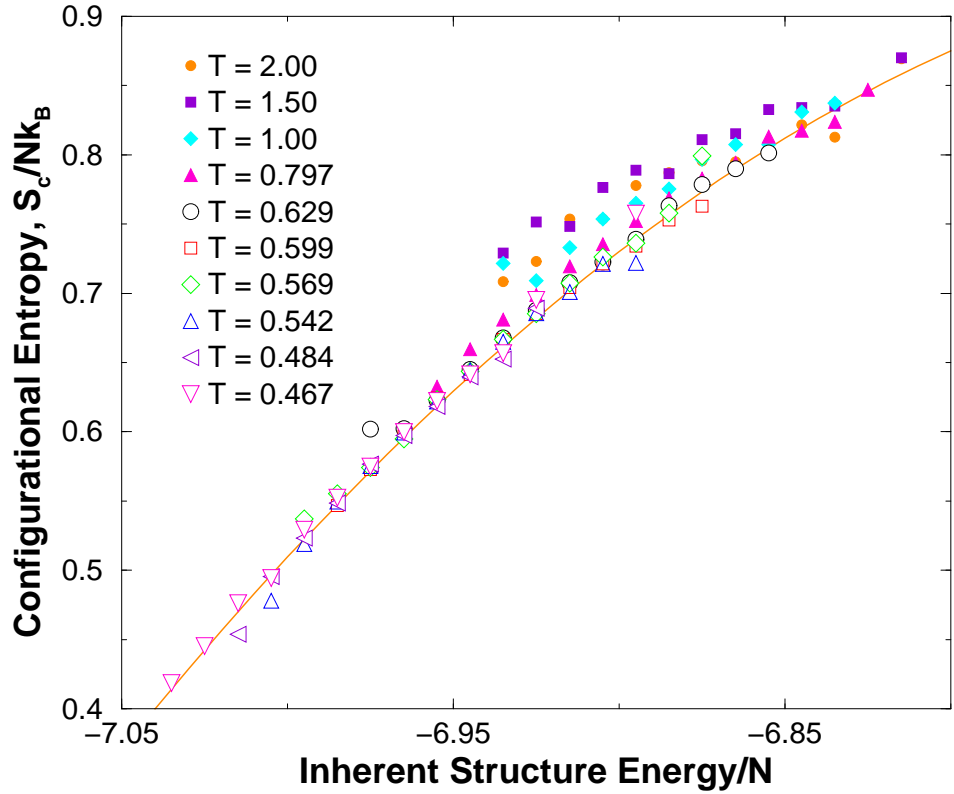}
\hfil}
\caption{Configurational entropy as a function of inherent structure energy 
per particle, obtained from a range of temperatures. The overlap of curves 
for $T < 0.8$ indicates that the harmonic approximation to the basin free 
energy is reasonable for $T < 0.8$. The solid line is a quadratic fit.}
\label{fig1} 
\end{figure} 

\begin{figure} 
\hbox to\hsize{\epsfxsize=0.8\hsize\hfil\epsfbox{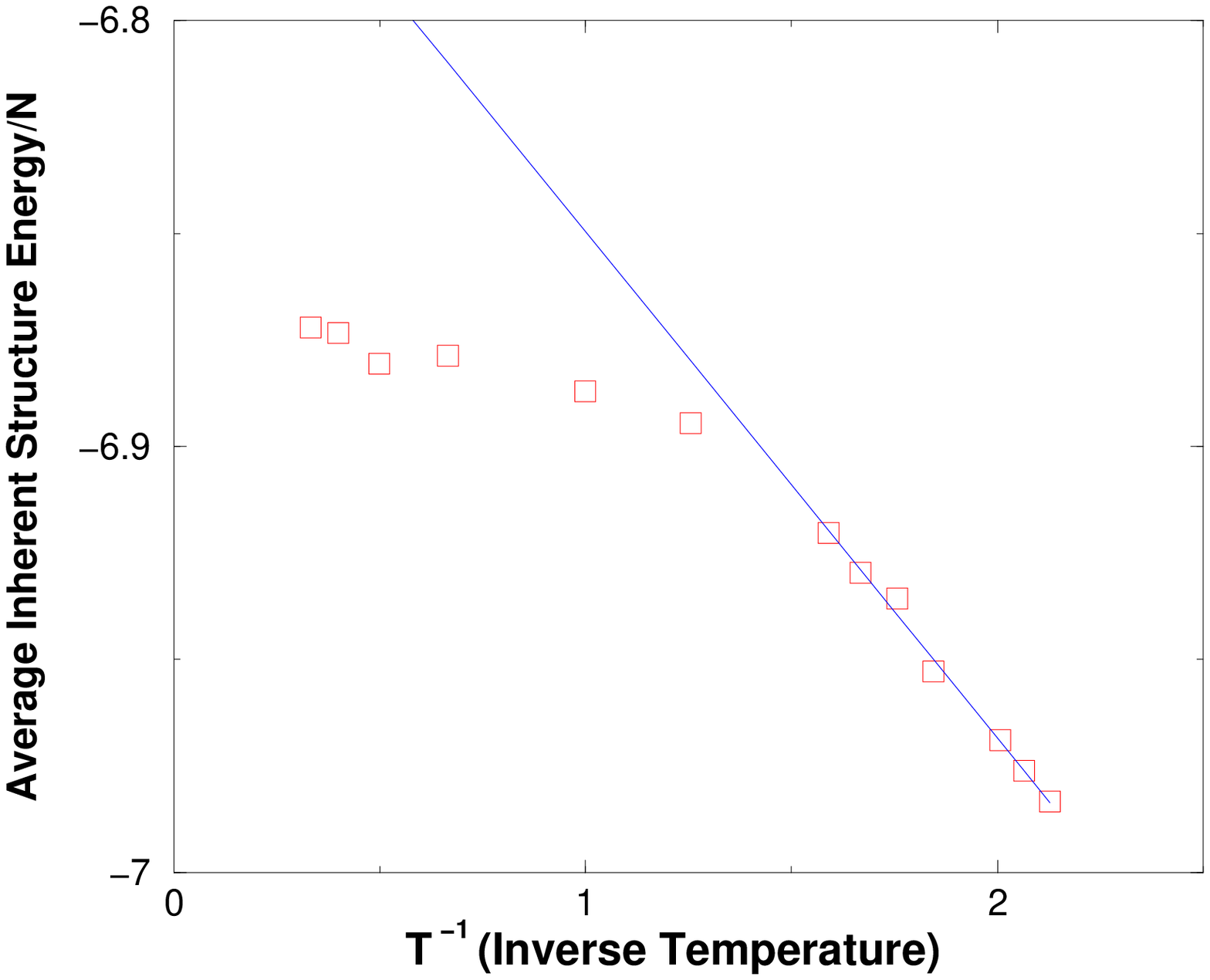}
\hfil}
\caption{Average inherent structure energy {\it vs.} inverse
temperature, showing that at low temperatures, the temperature
dependence is well-described by $T^{-1}$ while above $T = 0.8$ this 
dependence is not valid.}
\label{fig2} 
\end{figure} 

If the harmonic approximation to the basin free energy is accurate,
inversion of Eq. (\ref{scdos}), expressing $S_c(\Phi_\alpha)$ in terms
of $P(\Phi_\alpha, T)$, $Q_N(\rho,T)$ (or $A(\rho,T)$) and
$f_{basin}(\Phi_\alpha,T)$, for different temperatures $T$, should
result in curves that overlap with each other, as $S_c(\Phi_\alpha)$
is independent of $T$. Figure 1 shows the result of such inversion,
which indicates that below $T = 0.8$, the various $S_c$ curves do
overlap, while they do not at higher $T$. The procedure applied here
is similar to, but improves upon, the procedure of shifting
unnormalized $S_c$ curves adopted in \cite{fs,heuer}. Thus, Fig. 1
indicates that a harmonic approximation to the basin free energy is
not valid for temperatures higher than $T = 0.8$. The temperature
dependence of the average inherent structure energy $E_{IS}$, shown in
Fig. 2, is consistent with this conclusion. As discussed in \cite{heuer}, a
simple expectation for the T-dependence of the average inherent
structure energy in the harmonic regime is that $E_{IS} \sim 1/T$.
Figure 2 shows that such a T-dependence is indeed valid at low
temperatures, but breaks down for $T > 0.8$. However, this observation
must be viewed in conjunction with two other observations about the
topography of the inherent structure basins: (i) it has been
demonstrated recently \cite{schroeder} that the separation between
`vibrational' and `inter-basin' relaxation becomes reasonable for
temperatures close to and below the mode coupling $T_c$ ($ \sim 0.45$
for the model liquid studied here). (ii) The difference in the
potential energy of instantaneous configurations and the corresponding
inherent structures is nearly linear with a slope of $3/2$ for
temperatures as high as $T = 1.5$, as shown in Fig. 3. Such a linear
temperature dependence would normally be associated with harmonic
behavior, which in the present case is misleading.
\begin{figure} 
\hbox to\hsize{\epsfxsize=0.8\hsize\hfil\epsfbox{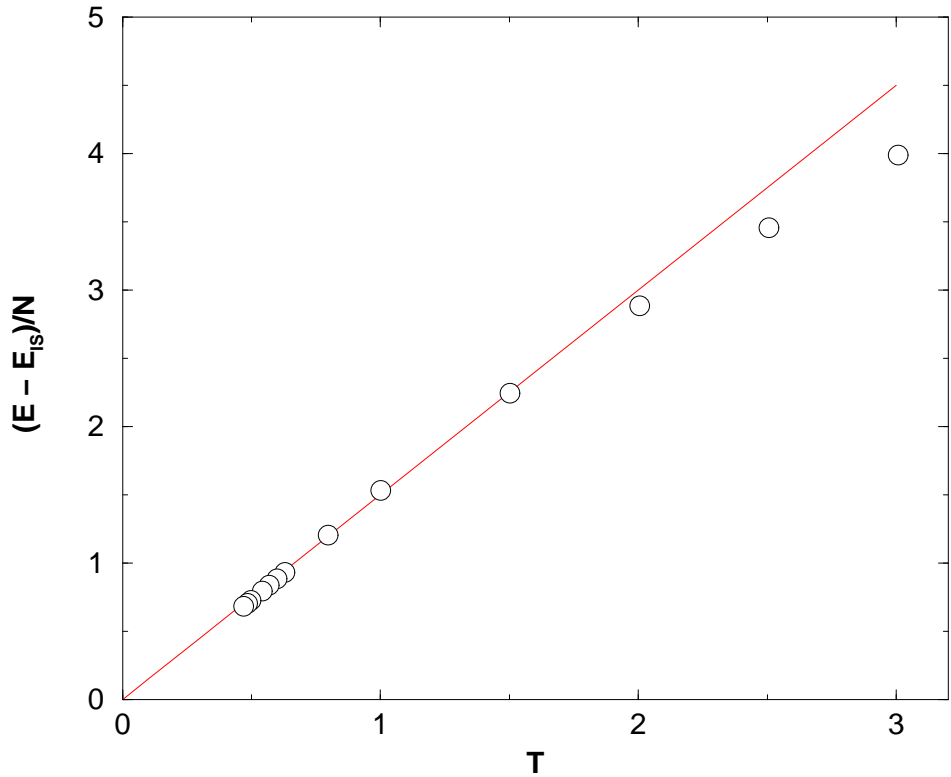}
\hfil}
\caption{Temperature dependence of the difference between the potential
energy of liquid configurations and their corresponding inherent structures,
which is linear with a slope of $3/2$ upto very high temperatures.}
\label{fig3} 
\end{figure} 

The total entropy of the liquid $S$ as well as the basin entropy
$S_{basin}$ are evaluated as a function of density and temperature
from the total and basin free energies. The configurational entropy
$S_c(\rho, T)$ and the ideal glass transition $T_{IG}(\rho)$ are then
given by,
\begin{eqnarray}
S_c(\rho, T) = S(\rho,T) - S_{basin} (\rho,T); S_c(\rho,T_{IG}(\rho)) = 0.
\end{eqnarray}

Figure 4 shows the configurational entropy so obtained as a function
of $T$. By extrapolation, based on the assumption that the potential
energy varied with temperature as $T^{3/5}$, the ideal glass
transition occurs at $T = 0.294$, in good agreement with estimates in
\cite{fs,parisi}.

\begin{figure} 
\hbox to\hsize{\epsfxsize=0.8\hsize\hfil\epsfbox{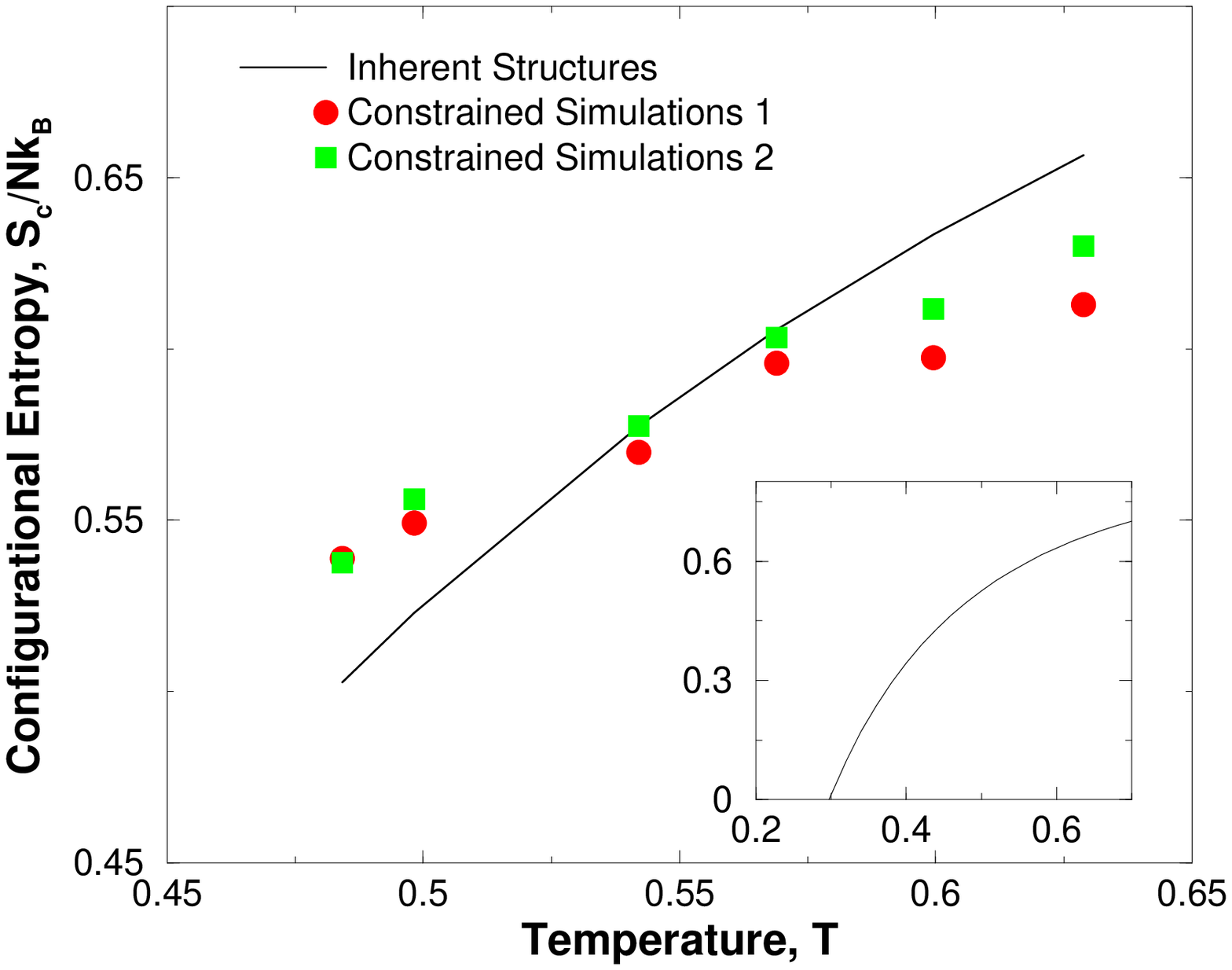}
\hfil}
\caption{Configurational entropy {\it vs.} temperature, obtained from
(a) Inherent structures (solid line), (b) constrained simulations,
where the constraint is applied to equilibrated liquid configurations
(filled circles, labeled `Constrained Simulations 1'), and (c)
constrained simulations, where the constraint is applied to inherent
structures (filled squares, labelled `Constrained Simulations 2'). Inset 
shows the extrapolation of $S_c$ which vanishes at $T = 0.294$.}
\label{fig4} 
\end{figure} 

\section{Constrained System Simulations}
\label{sec:section2}

	An alternate approach to defining the basin entropy, which has
been explored by Speedy\cite{speedy3} is to impose constraints on a
liquid to trap it in one of the basins it samples in equilibrium. A
related approach has also been studied in \cite{silvio}.  With
suitably chosen constraints, the calculated properties of the
constrained system allow the evaluation of the basin entropy. A
reasonable choice of constraint will restrict the system to a
physically meaningful set of configurations related to each other
without the need for configurational rearrangement. Further, such a
constrained system should behave reversibly. In this work, the
usefulness of one simple constraint is explored, by calculating the
configurational entropy for a set of six temperatures at a fixed
density of $1.2$, and compared with corresponding results from the
inherent structure calculations described above. It is found that the
constrained simulations result in comparable numbers for the
configurational entropy from the inherent structure results. 

	Ten sample configurations are chosen at each temperature, and
the Voronoi tessellation is performed for each configuration.  The
Voronoi cell of each given particle, and the corresponding geometric
neighbors, correspond to the {\it cage} a particle experiences at
short and intermediate time scales. A configurational rearrangement of
particles in the system will result in a restructuring of the Voronoi
tessellation as well. Thus, the constraint of restricting particles to
their Voronoi cells is an {\it a priori} reasonable choice. Hence, a
constraint is imposed which confines each particle to its Voronoi cell
during the Monte Carlo simulation from which the properties of this
constrained system are evaluated. Each Monte Carlo simulation
mentioned below is performed for $25, 000$ Monte Carlo steps. The
constrained system can be studied at any desired temperature; the
temperature of the simulation from which the reference configurations
are taken will be referred to as the fictive temperature where there
is need to distinguish these two temperatures. In order to estimate
the configurational entropy, we must evaluate the free energy of the
constrained system. This is done by thermodynamic
integration\cite{robinup,frenkel} from a reference system where each
particle experiences a harmonic potential around the initial
configuration (Einstein cystal). Considering a potential energy
function of the form,
\begin{equation} 
\Phi(\lambda, {\bf r}^N) = (1 - \lambda^2) (\Phi_{LJ}({\bf r}^N) + \Phi_c) + \lambda^2 C \sum_i ({\bf r}_i - {\bf r}_i^{0})^2 
\end{equation} 
where $\lambda$ is a tuning parameter that varies between $0$ and $1$,
$\Phi_{LJ}$ is the Lennard-Jones potential of the unconstrained
system, $\Phi_c$ is the constraining potential (which is zero if the
constraint is obeyed and infinity if it is not), the corresponding
free energy is given by
\begin{equation} 
A(\lambda,\rho,T) = -k_B T \log \left[\Lambda^{-3N} \int d{\bf r}^N \exp(-\beta \Phi(\lambda, {\bf r}^N))\right].
\label{cga}
\end{equation} 
The required free energy, $A(\lambda = 0, \rho, T)$ is related to that
of the Einstein crystal (which may be calculated straightforwardly), by 
\begin{equation} 
A(\lambda = 0, \rho, T) = A(\lambda = 1, \rho, T) - \int_0^1 {\partial A
\over \partial \lambda} ~d \lambda 
\end{equation}
where, from differentiating Eq. (\ref{cga}) with respect to $\lambda$, 
\begin{equation} 
{\partial A \over \partial \lambda}  =  -2 \lambda < \Phi_{LJ} - C \sum_i ({\bf r}_i - {\bf r}_i^{0})^2 >.
\end{equation} 
The required average in the above equation is calculated by performing Monte Carlo simulations for a set of $11$ $\lambda$ values. The values of $\partial A /\partial \lambda$ obtained are shown in Fig. 5. The free energy for $\lambda = 1$ is
\begin{equation} 
A(\lambda = 1, \rho, T) = 3 N k_B T \left[\log \Lambda - {1\over 2} \log({\pi \over \beta C})\right]. 
\end{equation}  
All the above calculations are also performed using the inherent structures 
corresponding to the equilibrated liquid configurations mentioned above. 
With the free energy of the liquid evaluated as described in the previous
section and the free energy of the constrained system obtained as described
here, the configurational entropy of the system is given by
\begin{equation} 
S_c/k_B = {A_{cs} \over N k_B T} - {A \over N k_B T} 
\end{equation} 
where $A_{cs}$ is the free energy of the constrained system. The
resulting configurational entropies are shown in Fig. 4. The $S_c$
values from the constrained system simulations are comparable with the
inherent structure results, but the agreement is moderate. In
particular, the constrained system results vary more weakly
with temperature. 

To verify that the chosen constraint is a reasonable one, the free
energies of the constrained system (for configurations from
equilibrium runs at (fictive) temperatures $T_f = 0.629, 0.484$) are
obtained independently at $T = 0.1$ and $T = 1.2$ from thermodynamic
integration with respect to the Einstein crystal. Simulations are also
performed for temperatures in between these two values, from which the
temperature dependence of the potential energy is obtained. Using
Eq. (\ref{Efit}), $A(T=0.1)/Nk_B T$ is calculated by integrating from
$T = 1.2$. The difference of the directly calculated value and the one
by integration is found to be $-0.0197$ for $T_f = 0.484$ and $.0247$
for $T_f = 0.629$. In other words, the constrained system appears to
be reversible within the margin of error represented by these numbers.
However, the discrepancy in the $S_c$ values between the constrained
system and the inherent structure estimates is of the same
order. Indeed, the discrepancy in the $S_c$ values for $T = 0.484$ and
$T = 0.629$ is roughly the same as the discrepancy in the free
energies above. It is likely that the sample of ten configurations
used here is too small to obtain more accurate values.

\begin{figure} 
\hbox to\hsize{\epsfxsize=0.8\hsize\hfil\epsfbox{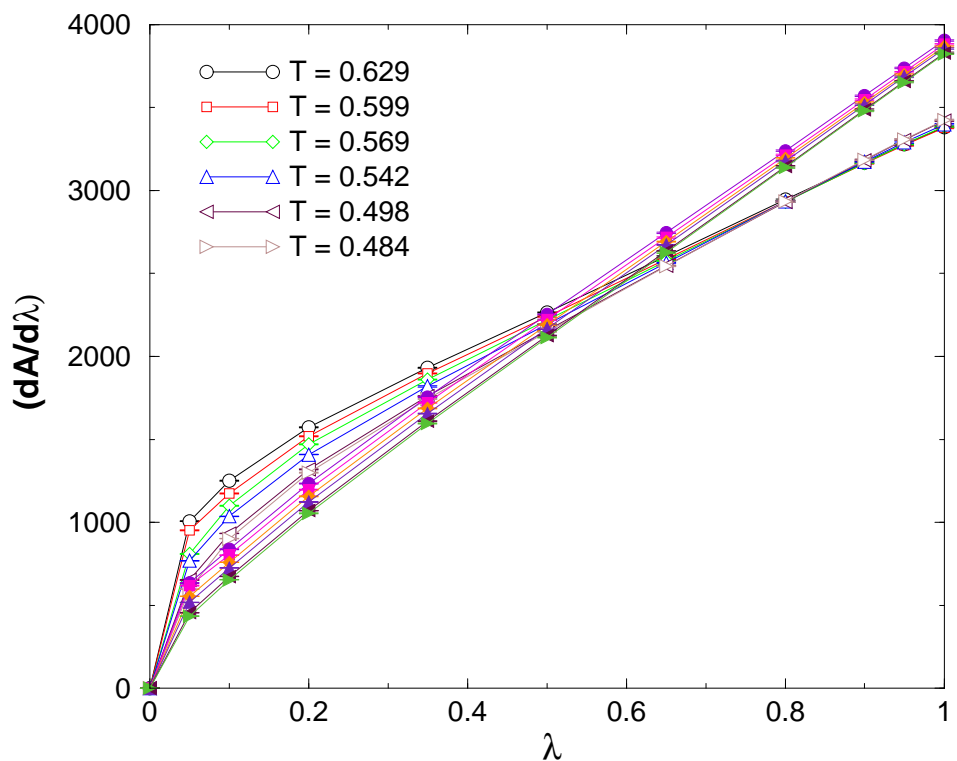}
\hfil}
\caption{Values of ${\partial A \over \partial \lambda}$ for temperatures 
as marked. The filled symbols represent values for systems constrained with 
respect to the corresponding inherent structures.}
\label{fig5} 
\end{figure} 

\section{Conclusions}
\label{sec:conclusions}

Configurational entropy is obtained for a binary mixture liquid from
analysis of inherent structures, and from estimation of the basin free
energy {\it via} constrained system simulations.  While the harmonic
approximation used in the inherent structure approach to evaluate the
basin free energy is in general questionable, the difficulty in the
constrained system approach is the proper choice of constraint. The
values for the configurational entropy obtained are comparable but
show only moderate agreement. Further tests for the accuracy of the
constrained system results, and more importantly, exploration of
improved constraining methods are desirable for making a more
stringent comparison of these two methods of calculating the
configurational entropy.

\section{Acknowledgments}
\label{acknowledgements}
Extremely useful discussions with Robin Speedy are gratefully acknowledged.

\end{document}